# Evaluation of the transmission dynamics of *Lawsonia intracellularis* in swine

Fernando L. Leite[1†*], Jason A. Galvis[2†], Dana Beckler[3], Steven McOrist[4]

[1]Boehringer Ingelheim Animal Health USA, 3239 Satellite Blvd Duluth, GA, USA

[2]Department of Population Health and Pathobiology, College of Veterinary Medicine, North Carolina State University, Raleigh, North Carolina, USA

[3]Gut Bugs Inc., Highway 120, Fergus Falls, Minnesota 56537, USA

[4]Scolexia P/L, 19 Norwood Crescent, Moonee Ponds, Victoria 3039, Australia

*Corresponding author: fernando.leite@boehringer-ingelheim.com

†These authors contributed equally to this work

## Abstract

The transmission dynamics of the common enteric pathogen *Lawsonia intracellularis* are not fully understood. To evaluate the transmission parameters of this pathogen, one and two conventional weaned pigs, were inoculated with a pure culture of *L. intracellularis*, then placed among respective groups (#1 and #2) of a sentinel cohort of pigs, 9 days later. The two experimentally exposed groups ($n$ = 31 pigs each) and a control group ($n$ = 5 pigs) were separately housed. Fecal shedding, seroconversion, oral fluid detection, and clinical signs were monitored throughout a 38-day exposure period following the introduction of the experimentally inoculated pigs. Transmission rates were estimated based on the number of infected pigs detected over time. Quantitative PCR first detected *L. intracellularis* in the two experimental groups 14 days post-exposure. By day 38, 63% of Group 1 and 86% of Group 2 sentinel pigs were infected. The estimated basic reproduction number ($R_0$) was 3.35 (95% CI: 1.62 - 7.03), and the transmission rate was 0.096 (0.046 - 0.2). These results demonstrate that introducing a single infectious source of *L. intracellularis* into a group of susceptible nursery age pigs is sufficient to sustain transmission over a prolonged period, resulting in infection in the majority of pigs within a five-week period. These findings underscore the ability of *L. intracellularis* to persist and propagate within a group, reinforcing the importance of detection and control measures.

## 1. Introduction

*Lawsonia intracellularis* is an obligately intracellular bacterium and the causative agent of proliferative enteropathy in pigs (McOrist et al., 1995). The infection and resulting enteric disease (clinical and subclinical) remain common among commercial and backyard pig farms worldwide (Vannucci et al., 2019). The natural infection occurs in an oral-fecal route, with subsequent proliferative mucosal lesions of the intestine occurring primarily in the ileocecal region (McOrist et al., 1993; Leite et al., 2019). Infection can lead to subclinical or clinical disease, causing a measurable reduction in animal performance, such as weight gain and feed conversion efficiency, with subsequent economic impact (Brandt et al., 2010; Gogolewski et al., 1991; Helm et al., 2021). *L. intracellularis* can persist in the environment of pig farms (such as pen floors and walkways) for two weeks or more (Collins et al., 2000), with possible carriage also occurring in fomites (such as farm boots) and the intestines of farm rodents and insect larvae (Backhans et al., 2013; McOrist et al., 2011).

Farm-raised pigs are normally weaned after 3 weeks of age and moved away from the breeding and farrowing area to be raised in a separate weaner or nursery area until approximately 10-weeks-of age, then moved to a separate finisher area until approximately 25 weeks of age, pigs may also be housed in a wean-to-finish building until reaching market age (Tucker et al., 2020). Previous studies using *Lawsonia*-specific serology and PCR have indicated the first infection of pigs can occur in the nursery phase of production, following natural waning of passive immunity (Hands et al., 2010; Hammer 2004). Recently, it has been shown that pigs may be infected with *L. intracellularis* at weaning at three weeks of age (Rodriguez-Vega et al., 2024). Fecal shedding is likely to start around seven days post infection and can persist intermittently for 12 weeks (Smith and McOrist 1997; Guedes and Gebhart, 2003). Although farm population-level epidemiology features of *L. intracellularis* have been studied (Bronsvoort et al., 2001; Stege et al., 2004), to the authors' knowledge, there are no previous studies to estimate transmission rate, basic reproductive number, or shedding rates necessary to further understand the disease dynamics and assist control strategies (Galvis et al., 2022a, 2022b).

Mathematical models, such as metapopulation models, can simulate infectious agent transmission among pig populations within different structures, connectivity, and conditions within pig farming systems (Andraud et al., 2023; Sicard et al., 2024). These models require robust data to approximate the natural course of the infection. For this study, we developed a strategy for challenge exposure to minimize confounder infections and assess the transmission dynamics of infected pigs introduced into a naïve population. We used the observed incidence to evaluate transmission dynamics of *L. intracellularis* between pigs at the pen level, providing information about the transmission rates, $R_0$, and shedding rates.

## 2. Materials and Methods

This study aimed to evaluate the transmission of *L. intracellularis* between infected and sentinel weaned pigs at the pen level over a five-week period. The study included 67 numerically tagged pigs of mixed Large White and Landrace breeds. All procedures and ethical considerations were approved under IACUC protocol 100887-RD, issued by Boehringer Ingelheim Animal Health, Duluth, GA, USA.

## 2.1. Study design, animals, and sample processing

*Sentinel pigs*

A total of 64 weaned pigs, approximately 21 days old, were selected from a single herd with no prior clinical history of *L. intracellularis* infections. Before inclusion in the study, all piglets and sows tested negative for *Lawsonia* fecal shedding by qPCR following methods previously described (Leite et al., 2021). The pigs were vaccinated before weaning (two weeks old) with commercial vaccines against *Mycoplasma hyopneumoniae* and porcine circovirus type 2. The pigs were subdivided into three groups and distributed into three separate pens, each in a different room within the same building (Figure 1). The two experimental groups comprised 30 and 29 sentinel pigs, respectively, while the control group contained five pigs. Group placement was determined using a random number generator. The experimental groups were housed in rectangular pens providing 0.24 $m^2$ per pig.

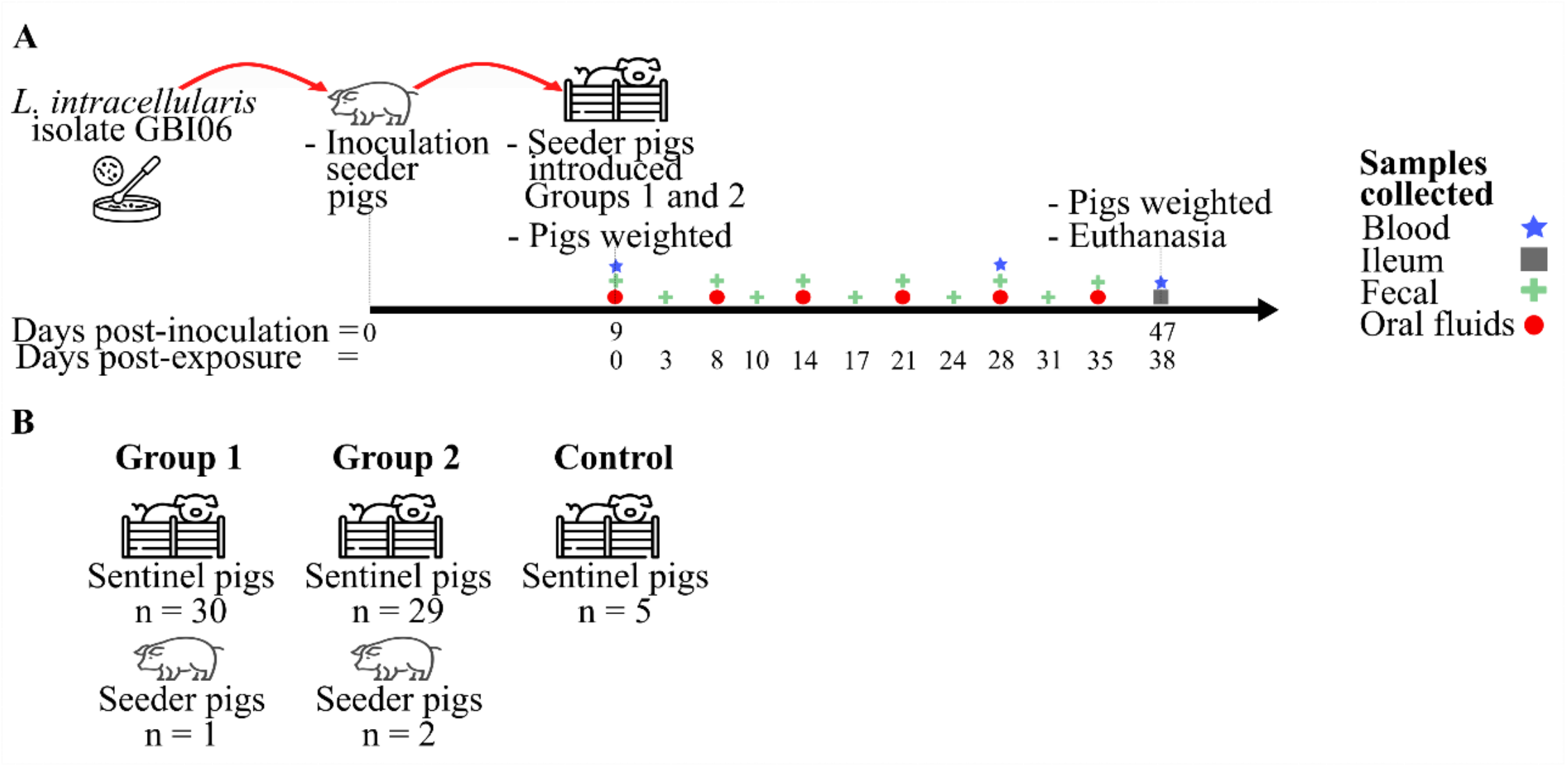

**Figure 1. Study design**. A) The timeline of experimental events and B) the distribution of sentinel and seeder pigs in the experimental and control groups.

*Inoculation of experimental seeder pigs*

Three additional pigs from the same source, also confirmed negative for *L. intracellularis*, were designated as seeder pigs 1, 2, and 3 and housed in a separate room to be experimentally inoculated and housed until the timing of exposure to sentinel group. These pigs were inoculated by oral gavage with $1.5 \times 10^7$ *L. intracellularis* organisms of isolate GBI06, following previously described methods (Guedes and Gebhart, 2003). The *L. intracellularis* isolate GBI06 was originally obtained from the intestine of a pig affected by proliferative enteropathy using standard *in vitro* co-culture methods (Guedes and Gebhart, 2003). The isolate underwent fewer than 10 passages, to assure maintenance of virulence and was stored at -80°C. Portions of the inoculum were tested to confirm the *L. intracellularis* content and purity from contaminants using standard protocols (Guedes and Gebhart, 2003). The challenge inoculated pigs were housed together and monitored daily for an initial nine days.

*Experimental procedure*

On day 1 of the experiment, one inoculated pig (seeder pig 1) was introduced into experimental Group 1 ($n$ = 30 pigs), and two other inoculated pigs (seeder pigs 2 and 3) were introduced into experimental Group 2 ($n$ = 29 pigs) (Figure 1). Each experimental group, therefore, consisted of a total of 31 pigs. No inoculated pigs were introduced into the control group ($n$ = 5).

All pigs were housed for 38 days (until approximately 10 weeks of age) following the introduction of inoculated seeder pigs and had *ad libitum* access to water and standard corn/soy-based nursery diets. Control and sentinel pigs were weighed, and blood samples collected at 0, 28, and 38 days post-exposure to seeder pigs (dpe). We used the Kruskal–Wallis test followed by Dunn's pairwise comparisons with Bonferroni adjustment to compare pig weights and daily weight gain between groups, with statistical significance set at $p < 0.05$. Each serum sample was analyzed using a commercial *L. intracellularis* ELISA assay as described previously (Jacobson et al., 2011). These sera collected at 38 dpe were further analyzed by an immunoperoxidase monolayer assay (IPMA) assay, as described previously (Guedes et al., 2002a). Fecal samples were collected from each pig on days 0, 3, 8, 10, 14, 17, 21, 24, 28, 31, 35 and 38 (Figure 1) and analyzed by quantitative PCR for *L. intracellularis* DNA following established methods (Leite et al., 2021). Fecal sampling was performed by collecting feces directly from the rectum with separate collection gloves for each animal. Oral fluids from each group were collected weekly using standard techniques and analyzed using the *L. intracellularis*-specific PCR as described above. Any clinical signs of all 67 pigs were monitored during each fecal sample collection, and fecal consistency was scored using a standardized system of 0 = normal, up to 3 = profuse watery diarrhea (Leite et al., 2018). At 38 dpe, all 67 pigs were humanely euthanized. A gross intestinal examination was performed, and a portion of the terminal ileum mucosa was collected for histological evaluation and immunohistochemical staining for intracellular *L. intracellularis*, as described previously (Guedes et al., 2002b).

## 2.2. Transmission parameters

We evaluated the epidemic growth curve and identified the growth rate (*r*), which quantifies the rate at which infected pigs increase during an early infection phase, using an exponential model (Kamvar et al., 2019):

$$\log (y) = r * t + b$$

where *y* represents the incidence of cases, *t* is time in days, *r* is the growth rate, and *b* is the intercept. A pig was defined as infected if it shed *L. intracellularis*. The basic reproduction number ($R_0$), which represents the average number of secondary infections by an infected pig over a stated period, was calculated using the Lotka–Euler equation (Wallinga and Lipsitch, 2007):

$$R_0 = \frac{1}{M(-r)}$$

where $M(-r)$ determines the distribution of the generation interval ($Tc$), and the shape of $M(-r)$ determines the appropriate relationship between the $R_0$ and the $r$. The generation interval $Tc$ represents the time between the infection of a primary case and its secondary cases. Since infected pigs can shed the bacteria as early as seven days post-infection (Guedes and Gebhart, 2003) and considering the homogeneous distribution of feces within the pen as the main route of infection, the exact relationship between new infectors and infectees across the experiment remains uncertain. Thus, we assumed $Tc$ as the average elapsed time between the introduction of the first infected pigs into the pens and the observation of the first secondary cases from both experimental groups. The generation interval distribution was modeled as a gamma distribution using the observed mean *Tc* from both groups and calculations were made using the $R_0$ R package (Obadia et al., 2012).

The transmission rate (β), which represents the rate per day infected pigs transmit *L. intracellularis* to susceptible pigs, was estimated using two methods. First, we derived β from the estimated $R_0$ and the recovery rate ($\gamma$) using the equation:

$$\beta = R_0 * \gamma$$

Since most pigs recover from clinical disease within four to six weeks (Guedes and Gebhart 2003; Collins, 2013), we used an average recovery period of five weeks (35 days) to estimate $\gamma$ as 1/35 days. Our second approach estimated β using an optimization approach to match the observed infection dynamics. Using the *optim* function in R (R Core Team, 2023), we implemented a quasi-Newton method ("L-BFGS-B") to adjust β within biologically reasonable bounds (0 to 0.2) (Byrd et al., 1995).

Finally, an Susceptible (S)-Infected (I) model was implemented with the estimated parameters, with the dynamics of S and I populations governed by:

$$\frac{dI}{dt} = \beta * S * I$$

The model outputs were compared to the observed accumulated incidence of infected pigs with *L. intracellularis*, both visually and by calculating the coefficient of determination ($R^2$) value to assess goodness-of-fit.

## 3. Results

The three seeder pigs (1, 2, 3) were successfully inoculated orally with a pure culture of *L. intracellularis* GBI06, leading to intestinal infection and persistent fecal shedding over the 47-day study period (Figure 2).

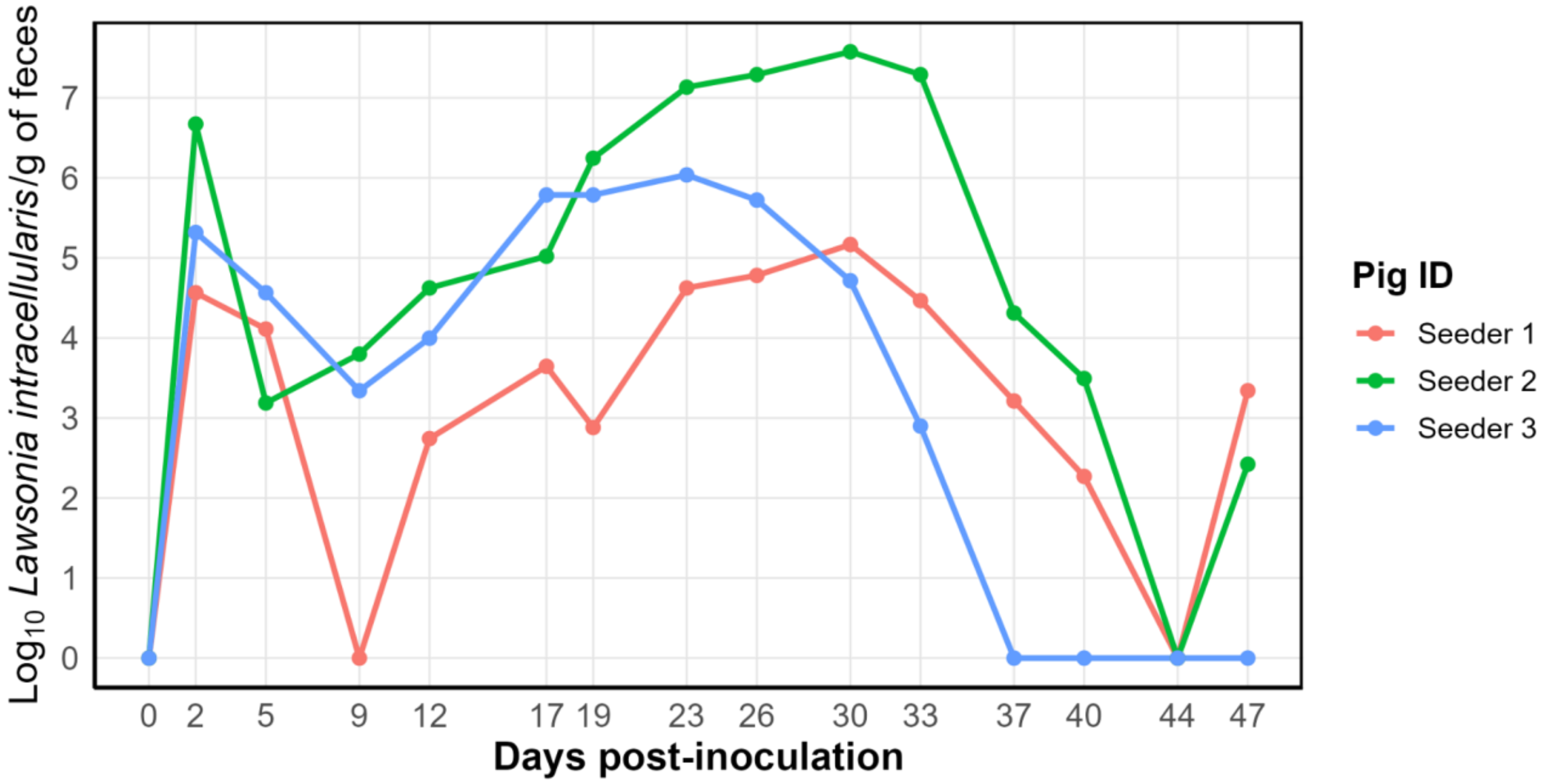

**Figure 2. Line plot showing the concentration of *L. intracellularis* shed per gram of feces over time in the three seeder pigs.** The y-axis is log-transformed, and the x-axis represents the timeline from inoculation to the trial endpoint.

The unchallenged sentinel and control pigs were all fecal PCR and seronegative on day 0 when the inoculated seeder pigs were introduced into the two sentinel pens.

Serologic assay of all pigs for *Lawsonia* antibody by ELISA and IPMA assay remained largely negative throughout. Only one pig in Group 2, which was seeder 2, was IPMA and ELISA positive at 38 dpe, while all the other pigs in the other groups remained negative.

PCR analysis of oral fluids showed two positive results at 14 and 21 dpe in Group 2, with Ct values of 33.9 and 34.54, respectively, while Group 1 showed one positive sample at 35 dpe with a Ct of 34.2.

Fecal shedding was first detected in one and four sentinel pigs from Groups 1 and 2 respectively, on 14 dpe. All samples from the control group were negative throughout.

During the study, *L. intracellularis* shedding was observed in 63% (19 of 30) of sentinel pigs in Group 1 and 86% (25 of 29) in Group 2 during the 38 dpe, with a median of 2.73 (interquartile range [IQR]: 2.27 - 3.19) $\log_{10}$ *L. intracellularis* per gram of feces across pigs from both groups. At 38 dpe, sentinel infected (positive) pigs in Group 1 were shedding a median of 3.19 (IQR: 2.96 - 5.56) $\log_{10}$ *L. intracellularis* per gram of feces, while Group 2 shed a median of 2.8 (IQR: 2.5 - 3.11) (Figure 3). Diarrhea (score three) was evident in approximately 10% of sentinel pigs in Group 1 at 8 dpe and 40% of the control Group (score 1 and 2). This level of diarrhea persisted in Group 1 for one week and then was noted again at 28 dpe with a score of two. Similarly, 14% of sentinel pigs in Group 2 showed diarrhea (score 2 and 3) at 21 dpe, but it was not persistent. Specific immunohistochemistry staining of the ileum revealed the presence of intracellular *L. intracellularis* in numerous epithelial cells within the crypts of the ileum in four sentinel pigs from Group 1, whereas no positive staining was observed in any pigs from Group 2. These four pigs were also actively shedding the bacteria, with fecal concentrations ranging from 5.94 to 7.47 $\log_{10}$ *L. intracellularis* per gram of feces.

The median pig weights at 30 days of age (time of seeder introduction) were 7.03 (IQR: 6.21 - 7.73) kg for experimental Group 1, 6.82 (IQR: 6.21 - 7.48) kg for experimental Group 2, and 6.8 (IQR: 6.2 - 7.5) kg for the control group, with no significant differences between groups (Kruskal-Wallis, $p > 0.05$). However, at 38 dpe, the median weights of pigs in experimental

Groups 1 and 2 were 29.8 (IQR: 27.6 - 31.4) kg and 28.6 (IQR: 26.6 - 30.6) kg, respectively, while the control group reached 34.5 (IQR: 33.3 - 37.1) kg (Figure 4). These results indicate a significantly reduced weight in both experimentally exposed groups compared to the control ($p < 0.05$).

The median daily weight gain in experimental Groups 1 and 2 was 0.46 (IQR: 0.42 - 0.49) kg/day and 0.45 (IQR: 0.41 - 0.47) kg/day, respectively, significantly lower than the 0.58 (IQR: 0.53 - 0.61) kg/day observed in the control group ($p < 0.05$).

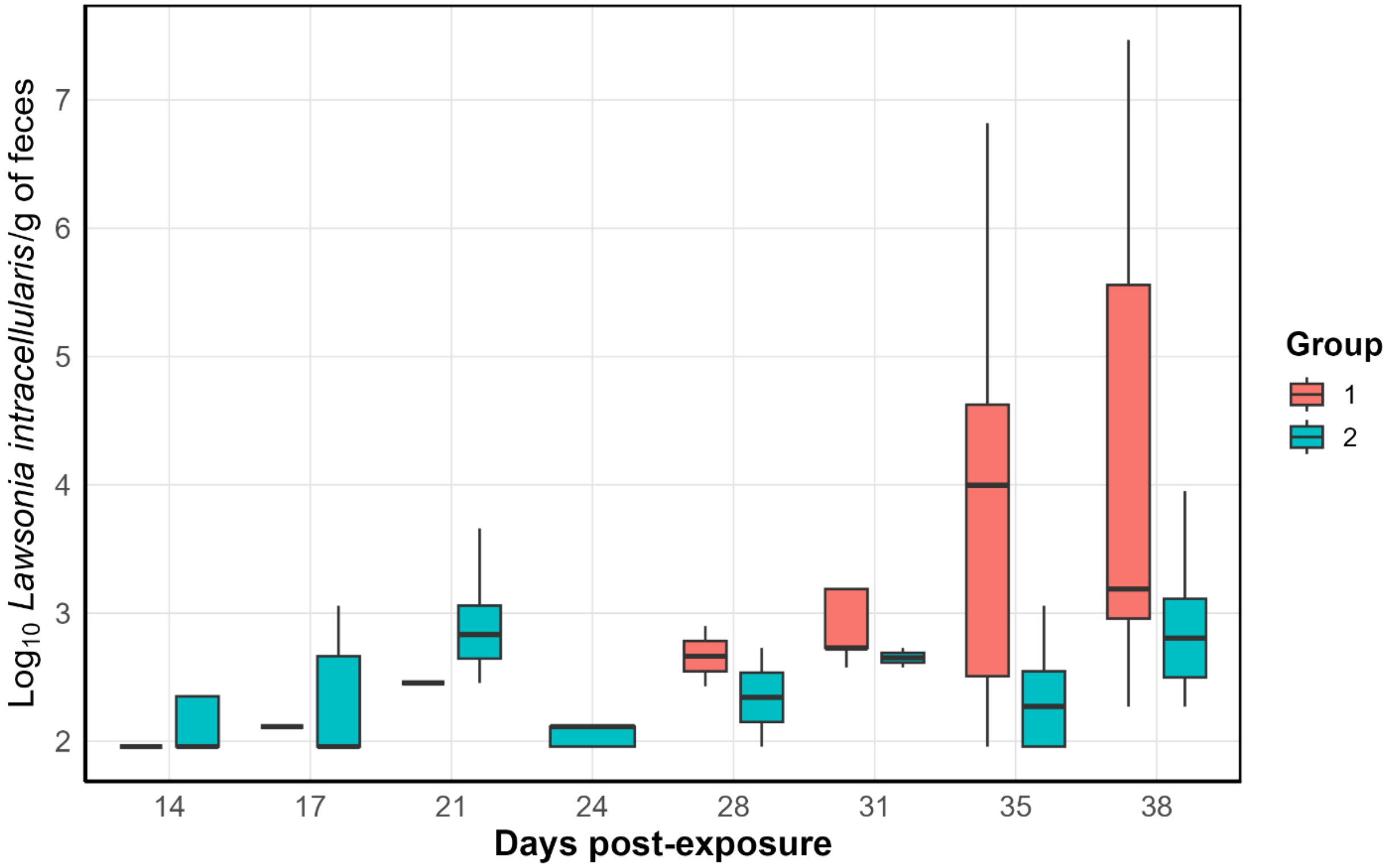

**Figure 3**. **Boxplot of the fecal shedding of *L. intracellularis* per gram of feces over 38 days post-exposure (dpe) from sentinel pigs from experimental groups one and two**. No shedding was observed in sentinel pigs from group one on day 24.

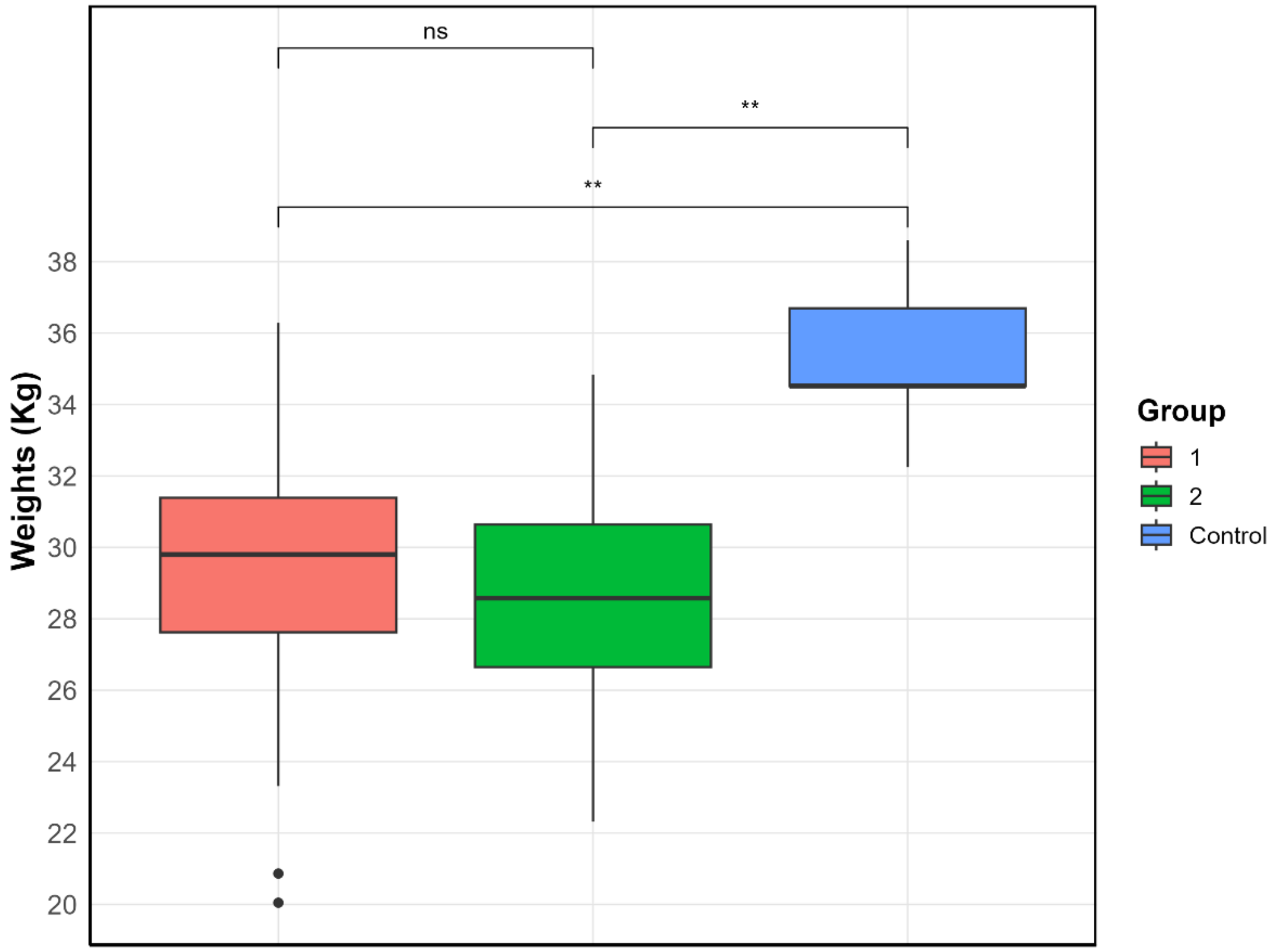

**Figure 4**. **Distribution of weight at 38 days post-exposure**. Horizontal lines above the boxplots indicate pairwise comparisons between groups using Dunn's pairwise comparison with Bonferroni adjustment. Significance levels are denoted as follows: *ns* = p > 0.05 and ** = p < 0.01.

Analysis of the rate of *L. intracellularis* infection indicated that it spread steadily in both exposed groups, with the cumulative incidence (including seeder pigs) reaching 64% (20/31) and 87% (27/31) by 38 dpe in Groups 1 and 2, respectively (Figure 5). The estimated daily epidemic growth rate for Group 1 was 0.087 (95% CI: 0.074–0.099), while for Group 2, it was 0.086 (95% CI: 0.061–0.11).

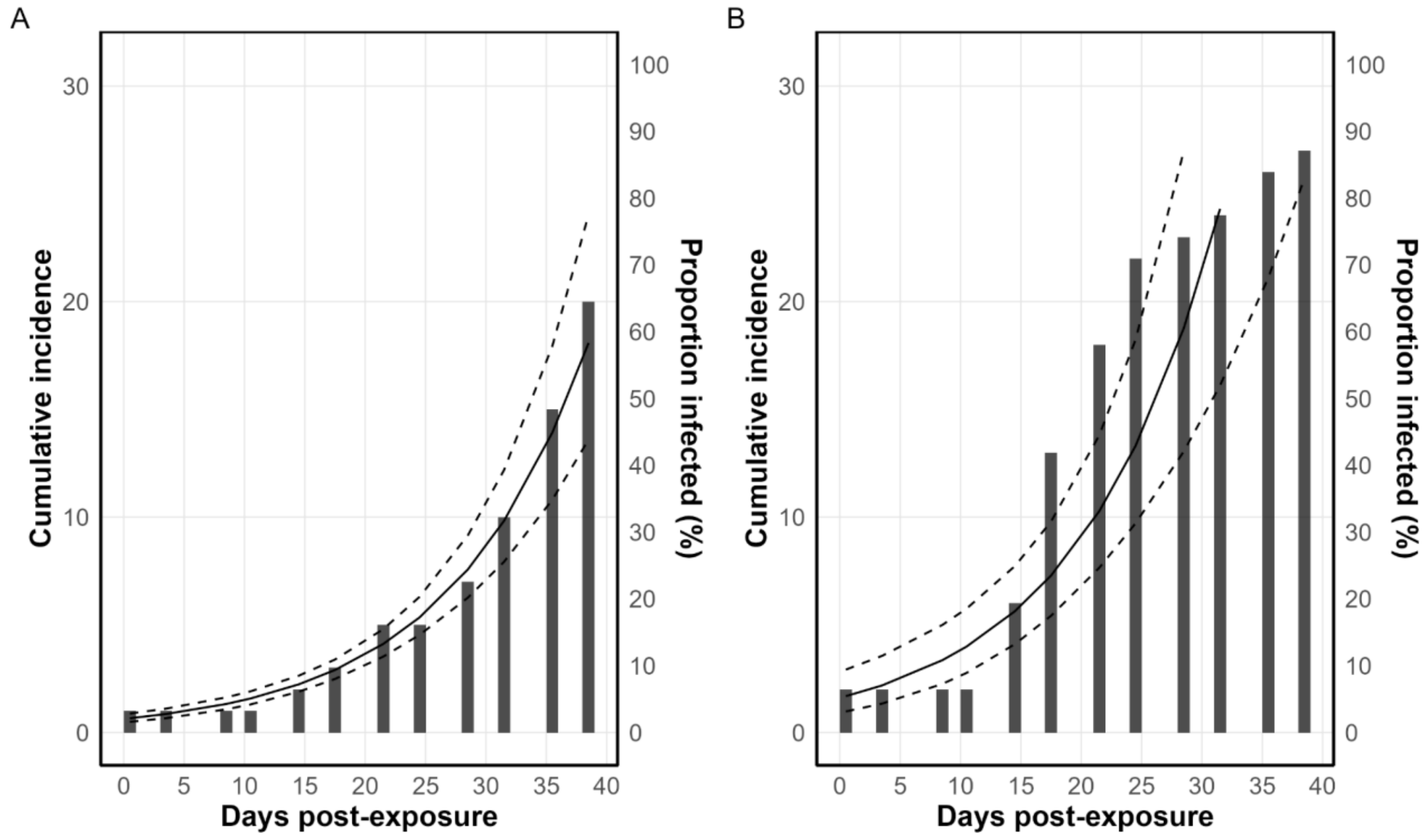

**Figure 5. Epidemic growth curve of the accumulated incidence of *L. intracellularis* fitted by an exponential linear model for seeders and sentinel infected pigs in experimental groups A) one and B) two.** The Y-axes shows the number of cases (absolutely and relatively), and the X-axis shows the number of days after starting the experiment. Solid lines indicate the fitted exponential models, and dashed lines show the 95% confidence intervals of the fit. The slope of each curve reflects the estimated growth rate.

The estimated generation interval had a mean of 14 days, and since both groups showed the first positive animals on the same day post-inoculation, we assumed a standard deviation of seven days, considering that pigs may start shedding *L. intracellularis* approximately seven days post-infection (Guedes and Gebhart, 2003). Group 2, with two challenged pigs introduced into the sentinel pen, created uncertainty regarding the exact relationship between infected and sentinel pigs. Therefore, we calculated $R_0$ only for Group 1, which showed an average of 3.35 (95% CI: 1.62–7.03). The transmission rate (β) estimates obtained using the two methods used were similar, with β = 0.096 (95% CI: 0.046–0.200) based on the estimated $R_0$ and β = 0.096 based on optimizing the observed infection dynamics. To assess the fit of the estimated transmission parameters, we compared the observed transmission dynamics with the median of simulated dynamics, obtaining a coefficient of determination of 0.83 for Group 1 and 0.79 for Group 2 (Figure 6).

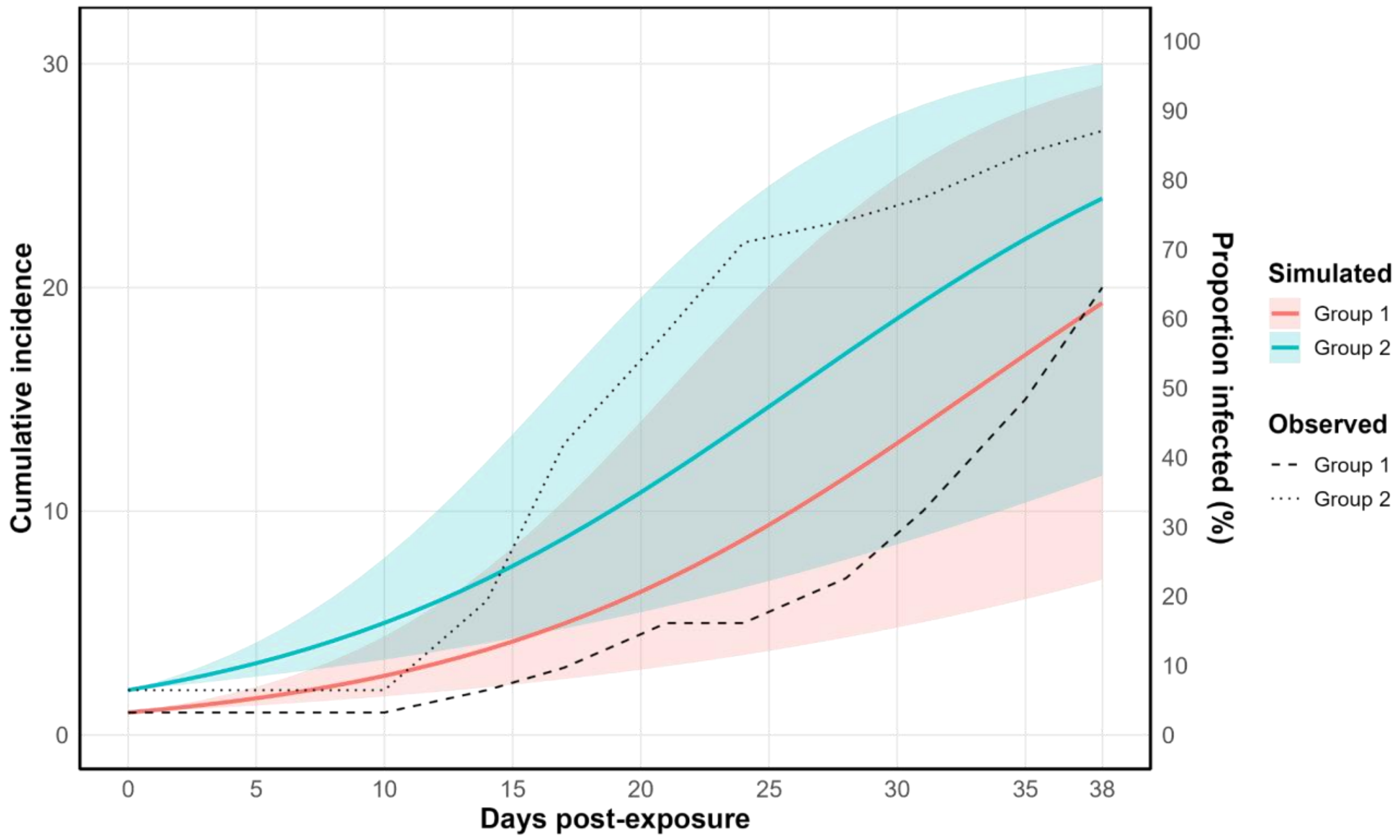

**Figure 6. Simulated incidence of *L. intracellularis*-infected pigs based on the estimated transmission rate (β) using a Susceptible-Infected model.** Dashed lines show the observed cumulative incidence. The solid lines indicate the estimated median simulated cumulative incidence, and shaded areas indicate the interquartile range.

## 4. Discussion

This study provides important insights into the transmission dynamics of *L. intracellularis* at the individual animal level within a pen of pigs. The estimated average basic reproduction number $R_0$ of 3.35, a transmission rate β of 0.096, and detectable shedding in 60–80% of infected pigs confirm that the infection can be sustained and efficiently disseminated among pigs under natural exposure conditions. The reduced growth rates in infected pigs with ongoing transmission in our study confirms numerous previous studies examining impaired growth rates and feed conversion in *Lawsonia*-infected pigs (Brandt et al., 2010; Gogolewski et al., 1991; Helm et al., 2021; Paradis et al., 2012). The more detailed characterization of *L. intracellularis* dynamics presented therefore offers a framework for further evaluating interventions, such as vaccination programs and enhanced biosecurity measures.

The estimated daily epidemic growth rates of 0.087 (95% CI: 0.074–0.099) for Group 1 and 0.086 (95% CI: 0.061–0.110) for Group 2 indicate a consistent rate of infection spread between the two experimental exposure scenarios. This similarity between the groups suggests that, despite numerical differences in the number of pigs used for initial exposure, the overall transmission potential of *L. intracellularis* under the experimental conditions was comparable. These epidemic growth rates also suggest a rapid initial transmission phase occurred in the pens under study, which may highlight the need for early detection and response strategies.

Notably, the consistency between the two transmission rate estimates (β = 0.096 based on $R_0$ and $y$ = 1/35 days, and β = 0.096 from model optimization) strengthens confidence in our findings. Furthermore, the goodness-of-fit, indicated by coefficients of determination of 0.83 for Group 1 and 0.79 for Group 2, demonstrates that the model captures the infection dynamics reasonably well, which was also observed by comparing the observed incidence falling within the credible intervals of the simulated transmission (Galvis et al., 2022a; Musa et al., 2020).

These values support the validity of the estimated parameters and reinforce the relevance of this approach for predicting disease spread and informing targeted interventions in commercial swine production settings (Andraud and Rose, 2020; Pittman Ratterree et al., 2024).

The transmission dynamics of enteric disease in pig populations naturally varies depending on the relevant microbial pathogen and virulence factors. In our study, pigs infected with *L. intracellularis* generated an average of 3.35 secondary cases over 35 days (infectious period according to our assumed recovery rate), with a modest transmission of 0.096 per pig per day, meaning each infected pig may transmit *Lawsonia* to about 0.1 susceptible pigs per day. Similarly, *Salmonella* Typhimurium also exhibits a relatively modest transmission, with an estimated β of 0.33 per pig per day and an $R_0$ around 1.9 (95% CI: 0.78–5.24) under field conditions (Correia-Gomes et al., 2014). In contrast, other enteric pathogens may spread more rapidly. Enterotoxigenic *Escherichia coli* has demonstrated an $R_0$ of 6.3 (95% CI: 1.89–21.48) in experimental weaned pig groups (Geenen et al., 2005), while porcine epidemic diarrhea (PED) virus has shown an $R_0$ of 5.39 in modeling studies of naïve populations (Makita and Yamamoto, 2019).

We found high rates of fecal shedding among the sentinel pigs in this study, with 63% and 86% of infected sentinel pigs shed detectable levels of the pathogen, with a median value of 2.73 (IQR: 2.27 - 3.19) $\log_{10}$ *L. intracellularis* per gram of feces, at least once during the observation period. This supports previous studies of the variable but often high shedding rates among exposed animals (Smith and McOrist 1997; Guedes and Gebhart, 2003). Alongside the known ability of *L. intracellularis* to persist in the environment, this supports the need for frequent cleaning of affected pig pens to minimize environmental contamination and limit further propagation of the pathogen between pens, barns, and farms (Collins et al., 2013).

Within detection limits, this study highlights significant limitations of monitoring *L. intracellularis* through the detection of pathogen DNA in oral fluids and serological assays. Although pigs were actively shedding the pathogen, all pigs except for one seeder pig were seronegative at the end of the study, and all pigs remained seronegative for *L. intracellularis* at 28 dpe. It is likely that pigs that were exposed and began shedding *L. intracellularis* at 14 dpe did not develop infection above a putative threshold needed for seroconversion. Very low levels of seroconversion ranging from 12.5% and 37.5% has been observed in pigs experimentally directly challenged with doses between 4 and 5 $\log_{10}$ *L. intracellularis* (Paradis et al., 2012). While oral fluids have recently been described to detect *Lawsonia* by PCR (Eddicks et al., 2025), in this study, a pen with positive pigs yielded mostly negative results. This highlights the need to further characterize this sample type and better determine its sensitivity. A consequence of these findings is that monitoring for *Lawsonia* should likely be focused on fecal samples rather than blood or oral fluid samples. Given that a single positive pig shedding at low levels can still comfortably lead to transmission among many other contact pigs means that eradication efforts for *Lawsonia* are likely to be difficult to monitor and accomplish. Future studies should consider trials to evaluate vaccination and sources of infection, ultimately contributing to preventing the continuous circulation of *L. intracellularis* among pigs within the farm.

**Limitations and further remarks**

While the exponential growth model is a valuable tool for estimating transmission dynamics, it has important limitations (Kamvar et al., 2019). Its accuracy depends on the selected time window, which may bias estimates if it includes the peak or saturation phase of the epidemic. The method also assumes purely exponential growth and does not account for observation error, stochastic variation, or sub-exponential patterns. Although our experiment was controlled and sampling was consistent over 38 days, the estimated growth rates should still be interpreted with caution. Another limitation relates to our assumption of seven days standard deviation for the generation interval, which was not directly measured. This value was chosen to reflect the expected variability in infection timing due to the indirect fecal-oral transmission pathway, differences in individual exposure, and uncertainty in the precise timing of secondary infections. Given that pigs can begin shedding *L. intracellularis* as early as seven days post-infection (Guedes and Gebhart, 2003), and that environmental transmission likely spread exposure over multiple days, a standard deviation of seven days appears biologically reasonable. Moreover, the consistency between our model outputs and observed data provides additional support for this assumption. A limitation of the ability to calculate $R_0$ was that the number of introduced infected pigs to sentinel groups was unequal, therefore, the calculation of $R_0$ was performed with data from Group 1 only. Also, some pigs in the control group developed diarrhea, which suggests that a factor other than *L. intracellularis* might have been present and could have influenced the observed clinical signs. The frequency of sampling was approximately every three days and not daily, which could lead to less precision in determining infection states. An increase in sampling frequency, however, would drastically increase the cost of the study. Finally, we interpreted fecal shedding as a proxy for infection, but this event may also reflect the transient passage of bacteria acquired from the environment, rather than true infection. To our knowledge, this possibility has not been confirmed for *L. intracellularis,* and it is known that infection can occur as soon as 12 hours post-inoculation (Boutrup et al., 2010). However, in this study, infection was confirmed in four sentinel pigs from Group 1 through immunohistochemistry, which demonstrated the presence of *L. intracellularis* within the ileal epithelium, indicating active intracellular infection rather than environmental contamination. Notably, this number aligns closely with the estimated $R_0$ of 3.35, supporting the biological plausibility of the transmission dynamics observed in this study.

## 5. Conclusion

*L. intracellularis*, when introduced via seeder pigs into susceptible weaned pig populations, infected up to 86% of the pigs and significantly impacted the weight gain of infected groups. We evaluated the transmission dynamics of *L. intracellularis,* in which the average $R_0$ was estimated as 3.35 (95% CI: 1.62–7.03) and the per day transmission rate (β) of 0.096 (95% CI: 0.046–0.200), indicating efficient and steady within-pen dissemination. This study also highlights limitations in detecting *Lawsonia* infection using oral fluids and serological assays.

## Authorship contribution statement

FLL, JAG, DB, and SM conceived the study, participated in the study's design, and coordinated the data collection. FLL and JAG conducted data processing, cleaning, and designing data analysis. JAG designed the computational analysis. FLL, JAG, DB, and SM wrote and edited the manuscript, and FLL secured the funding.

## Declaration of Interest

All authors confirm that there are no conflicts of interest to declare.

## Data Availability

The data that support the findings of this study are publicly available in the Supplementary Material.